\documentclass[useAMS,usenatbib,referee]{biom}
\usepackage{amsmath}
\usepackage[figuresright]{rotating}
\usepackage{multirow}
\usepackage{cite}
\usepackage{hhline}
\usepackage{natbib}
\usepackage{bm}
\usepackage{mathrsfs}
\usepackage{amssymb}
\usepackage{xcolor}

\onecolumn

\title[Latent Multivariate Log-Gamma Models for High-Dimensional Multi-Type Responses]{Latent Multivariate Log-Gamma Models for High-Dimensional Multi-Type Responses with Application to Daily Fine Particulate Matter and Mortality Counts}

\author{Zhixing Xu$^{1*}$\email{zhixing.xu@stat.fsu.edu},
        Jonathan R.Bradley$^{1}$, Debajyoti Sinha$^{1}$\\
   $^{1}$Department of Statistics, Florida State University, Tallahassee, Florida 32306, U. S. A.\\}

\begin{document}

\begin{abstract}
Tracking and estimating Daily Fine Particulate Matter (PM2.5) is very  important as it has been shown that PM2.5 is directly related to mortality related to lungs, cardiovascular system, and stroke. That is, high values of PM2.5 constitute a public health problem in the US, and it is important that we precisely estimate PM2.5 to aid in public policy decisions. Thus, we propose a Bayesian hierarchical model for high-dimensional ``multi-type" responses. By ``multi-type" responses we mean a collection of correlated responses that have different distributional assumptions (e.g., continuous skewed observations, and count-valued observations). The Centers for Disease Control and Prevention (CDC) database provides counts of mortalities related to PM2.5 and daily averaged PM2.5 which are both treated as responses in our analysis. Our model capitalizes on the shared conjugate structure between the Weibull (to model PM2.5), Poisson (to model diseases mortalities), and multivariate log-gamma distributions, and we use dimension reduction to aid with computation. Our model can also be used to improve the precision of estimates and estimate values at undisclosed/missing counties. We provide a simulation study to illustrate the performance of the model, and give an in-depth analysis of the CDC dataset.
\end{abstract}


\maketitle

\section{Introduction}
%what problem we want to solve, the motivation
The National Academy of Sciences has consistently labeled Daily Fine Particulate Matter (PM2.5) as an important quantity to monitor to aid in the assessment of US public health \citep{burnett2018global}. This is partially due to the fact that PM2.5 is highly correlated with incidence/mortality of several diseases \citep{laden2000association,schwartz2000fine,valavanidis2008airborne}. For example, the findings in \citet{turner2011long} strengthened previous  evidence that increases of concentrations of PM2.5 are associated with increases in lung cancer mortality among ``never-smokers".  \citet{anderson2012clearing} reviews several studies on the effects of particulate matter air pollution on human health including stroke. 
\citet{brook2010particulate} concludes that long-term exposure to PM2.5 will increase the mortality due to cardiovascular problems.  

These relationships imply that there is an opportunity to leverage the dependence between PM2.5 and mortality counts to improve the precision of the estimates of both PM2.5 and mortality. Several federal agencies provide data on PM2.5, including the centers for Disease Control and Prevention (CDC). Each year the CDC provides hundreds of summary statistics regarding cancer incidence, mortality, and risk and screening behaviors on US counties (https://www.cdc.gov/). Several authors have used spatial statistical models to analyze these data \citep{clarke1996epidemiology,chaput2002spatial,eisen2007need,mollalo2015geographic,tarr2018geogenomic}. However, spatial statistical models are defined for a single type of response (e.g., either all continuous or all counts responses) instead of multi-type responses (e.g., one continuous and another count-valued) such as continuous response PM2.5 and mortality counts. Thus, our primary goal is to model PM2.5 and mortality counts using a statistical model that leverages spatial dependence as well as dependence between PM2.5 and mortality at each location. 

%first contribution: multi-type data
There is a growing literature for methods to analyze correlated multi-type responses. For example, there exists regression trees, copulas, and machine learning type algorithms for high-dimensional multi-type responses \citep[e.g., see][]{dobra2011copula,liu2009nonparanormal,xue2012regularized,liu2012high}. However, parametric models for this setting has been given considerably less attention and there are only a few examples of this type of joint modeling in the parametric setting \citep{sammel1997latent, yang2014general, wu2015bayesian}. Only \citet{wu2015bayesian} uses a Bayesian approach, and only for multi-type data restricted to Binomial/Poisson responses, unlike Weibull/Poisson responses for our motivating CDC dataset. To emphasize that we are capitalizing on the dependence between response types, we refer to our model as the joint Weibull and Poisson (WAP) model. 

%second contribution: log-gamma process as priors 
We jointly analyze the CDC's PM2.5 responses and mortality counts by intricately combining existing models in the literature. In particular, we use the multivariate log-gamma distribution \citep{bradley2018bayesian} to obtain easy to sample from conjugate updates within a collapsed Gibbs sampler for our multi-type model. Directly sampling from a conjugate full-conditional distribution is particularly important because this allows one to avoid tuning parameters and defining proposal distributions in Markov Chain Monte Carlo (MCMC) algorithms. 
This choice to modeling joint random effects in Weibull and Poisson data an important contribution. In particular, the use of the MLG distribution has been used to accurately model continuous skewed \citep{hu2018bayesian} data and count data \citep{bradley2018bayesian}, but has not been used to jointly model counts and continuous observations.

%third contribution: reduce rank
An important goal of this paper is to introduce a model that allows for computationally efficient Bayesian inference of large datasets similar inside of the motivating CDC dataset. In particular, we allow a reduced rank expression of spatially co-varying terms. Reduced rank spatial models have been shown to have high predictive accuracy and be computationally efficient \citep{wikle1999dimension,cressie2006spatial,shi2007global,banerjee2008gaussian,cressie2008fixed,finley2009improving,katzfuss2011spatio,bradley2015multivariate,heaton2018case}. To our knowledge, no such reduced rank methodology has been applied to correlated multi-type responses. This is especially notable because reduced rank uni-type multivariate spatial models have been shown to work well, but can be sensitive to the choice of the number of basis functions for the spatial pattern \citep{bradley2011selection, stein2014limitations, bradley2019hierarchical}. 

% Robustness to zero counts
In addition to efficient computation, our WAP model based analysis is also robust to inflation of zero counts. It is well-known that the inflation of zero responses/counts may lead to biased estimation for some common models of counts, and consequently, a rich literature is available for various zero-inflated models \citep[e.g., ][, for a discussion]{sellers2016flexible}. Via simulation studies, we show that our model performs well even when there are inflated numbers of zeros mortality counts. This property is particularly important because our motivating CDC database contains a moderate amount of observed zero mortality counts. 

Another motivation for the  proposed model is that it can easily be adapted to other studies; hence, the model is of independent interest. Skewed data with correlated counts arise in several disciplines \citep[e.g., daily sunlight (hours) may be correlated with melanoma incidences, ][]{leiter2008epidemiology}. Furthermore, our approach is computationally feasible, and as a result, the WAP model can be applied to other studies with correlated continuous and count valued observations, and with similar computational challenges.

% the structure of the rest paper
In Section~\ref{model and method}, we review the multivariate log-gamma distribution and introduce the WAP model to jointly model high-dimensional multi-type survival responses. In Section~\ref{simulation study}, we conduct a simulation study to compare the performance of WAP against competing models that ignore the dependence between Weibull and Poisson data. In Section~\ref{data analysis}, we use the WAP model to analyze the the aforementioned CDC dataset and illustrate the performance of the model through simulations. We provide a conclusion in Section~\ref{discussion}. All proofs and software are given in an Appendix for ease of exposition.

\section{Model and Method \label{model and method}}
\subsection{A Uni-Type Model for Weibull Responses \label{Weibull model}}
Consider a dataset organized into the $n_c$-dimensional vector $\bm t =\{t(A_1),t(A_2),\dots,t(A_{n_c})\}'$ consisting of a Weibull random variable. That is, $t(A_i) \stackrel{ind}\sim Weibull(\rho_i, b_i)$ for $i =1,\dots,n_c$, where we let the scale parameter $\bm b=(b_1,b_2,\dots,b_{n_c})'$ be parameterized as $b_i=exp(-Y_c(A_i))$ for $i=1,\dots,n_c$,  and the subscript ``c" indicates a continuous response.  Let $A_i\in D$ denote the $i$-th region $i=1,\dots,n_c$, where $D$ denotes the spatial domain of interest (e.g., the U.S.). For example, $A_i$ might represent a county, state, etc.. The areal units within the domain $D$ are disjoint, that is $\cup_{i=1}^{n} A_i\subset D$ and $A_i \ne A_j$ for $i\ne j$. The density function of the Weibull distribution we used here is
\begin{equation*}
f(t(A_i)|\rho(A_i),Y_c(A_i))=\rho(A_i) t(A_i)^{\rho(A_i)-1}exp[Y_c(A_i)-t(A_i)^{\rho(A_i)}exp\{Y_c(A_i)\}]
\end{equation*}
for $i =1,\dots,n_c$.

The primary goal of our analysis is to predict $Y_c(A^*)$, where $A^*\in D$ but $Y_c(A^*)$ is not observed. That is $A^*$ represents a county that does not have continuous response available disclosed/observed. It is straightforward to allow for both covariate and spatial effects to aid in estimating $Y_c(A^*)$. Specifically, let
\begin{equation*}
Y_c(A)=\bm x_c(A)'\bm \beta_c+\bm{\psi}_c (A)'\bm\eta_c+\gamma_c(A);\ A\in D,
\end{equation*}
where $\bm x_c(A)$ is a $p_c$-dimensional vector of known covariates, and $\bm \beta_c$ is a $p_c$-dimensional parameter vector of the covariate effects. The $r$-dimensional random vector $\bm \eta_c$ represents the spatial random effects. The $r$-dimensional term $\bm{\psi}_c(A)$ is pre-specified and can be any class of areal basis function (e.g., aggregations of thin plate spline basis function and bisquare basis function see \citet{supbradley2018bayesian}). We give example choices of $\bm{\psi}_c(\cdot)$ in Sections~\ref{simulation study} and \ref{data analysis}.

The term $\bm\psi_c(A)'\bm\eta_c$ is a spatial basis function expansion, which has become a standard tool in modern spatial analysis \citep{wikle2010general}. The random error $\gamma_c(A)$ is assumed to capture unknown spatial error not accommodated by $\bm{x}_c(A)'\bm{\beta}_c+\bm{\psi}_c(A)'\bm{\eta}_c$. In other words, $\gamma_c (A)$ captures fine-scale variability that is smoothed across by $\bm{x}_c(A)'\bm{\beta}_c+\bm{\psi}_c(A)'\bm{\eta}_c$. This follows the standard decomposition of a spatial process into large scale (i.e., $\bm x_c(A)'\bm\beta_c$), small scale (i.e., $\bm \psi_c(A)'\bm \eta_c$) and fine-scale (i.e., $\gamma_c(A)$) variability \citep{cressie2015statistics}. In Appendix C, we give a complete expression of a model for Weibull responses. This includes prior and hyper prior specifications, which are chosen to be conjugate \citep{diaconis1979conjugate,bradley2018bayesian}. In particular, $\bm\beta$ and $\bm\eta_c$ follow an MLG distribution (see Appendix A for more details). Classic implementation of (\ref{mlg}) and (\ref{naturalPc}) assumes that $\bm\beta$ and $\bm\eta_c$ are Gaussian random vectors \citep{bradley2015multivariate}. The MLG distribution has some advantages over the Gaussian distribution. Specifically, the MLG distribution leads to easy to sample from full conditional distributions (see Appendix B), and can be specified arbitrarily close to the Gaussian distribution \citep{bradley2018bayesian}.

\subsection{Multivariate Log-Gamma Distribution}
We assume that random effects in our model are distributed according to the MLG distribution. Thus, in this section we give a short review on the relevant details on the multivariate log-Gamma distribution.  Let $\bm w=(w_1,w_2,\dots,w_m)'$ be an $m$-dimensional random vector of $m$ mutually independent log-gamma random variables  $w_i=log(\gamma_i)$, where $\gamma_i$ is a gamma random variable with shape $\alpha_i>0$ and rate $\kappa_i>0$. Then, the multivariate log-gamma random variable $\bm{q}\sim MLG(\bm c,\bm V,\bm\alpha,\bm\kappa)$ is defined as
\begin{equation}\label{mlg}
\bm{q}=\bm{c}+\bm{Vw},
\end{equation}
where $\bm c$ is a $m$-dimensional vector, $\bm V$ is a lower-triangular $m\times m$ invertible matrix, $\bm{\alpha}=(\alpha_1,\dots,\alpha_m)'$ and $\bm{\kappa}=(\kappa_1,\dots,\kappa_m)'$. The probability density function (pdf) of the $m$-dimensional random vector $\bm q$ is, 
\begin{equation*}
f(\bm q|\bm{c,V,\alpha,\kappa})=\frac{1}{det(\bm V'\bm V)^{\frac{1}{2}}}\bigg(\prod^m_{i=1}\frac{\kappa_i^{\alpha_i}}{\Gamma(\alpha_i)}\bigg) exp\bigg\{\bm{\alpha}'\bm {V^{-1}(q-c)}-\bm \kappa'exp\big[\bm V^{-1}(\bm q-\bm c)\big] \bigg\}.
\end{equation*}
We use the MLG distribution to be priors of random variables in Weibull response model except $\bm \rho$. A Gamma prior is assigned for each $\rho_i$ and we use Metropolis-Hasting algorithm to update it. The main advantage of the MLG distribution is that the likelihood of $\bm t$ and Poisson counts (with log link) have a double exponential form similar to that of the MLG distribution. This can be exploited when implementing a Gibbs sampler (see Appendix B). Further details on the properties of the MLG distribution is provided in Appendix C. 

\subsection{Joint Modelling Weibull and Poisson (WAP) Responses}
In practice datasets consisting of multiple types are not mutually independent. Consider a dataset $\mathscr{D}=\{\mathscr{D}_c,\mathscr{D}_d\}$ consisting of two different types of responses: a dataset $\mathscr{D}_c$ consisting of continuous responses, and a dataset $\mathscr{D}_d$ consisting of count-valued observations. The continuous response is made-up of responses distributed according to the Weibull distribution, 
 $\mathscr{D}_c=\{t(A_i): A_i\in D,i=1,\dots,n\}$. The count-valued dataset is defined as $\mathscr{D}_d=\{Z(A_i): A_i\in D, i=1,\dots,n\}$, where $Z(A_i)$ is assumed to be Poisson. We introduced a shared spatial basis function expansion to model the dependence between these two different types of responses. That is, the WAP model makes the following assumption: 
\begin{equation}\label{naturalPc}
Y_c(A)=\bm{x}_c(A)'\bm \beta_c+\bm \psi_c(A)'\bm \eta+\bm \psi_c(A)'\bm \eta_c+\gamma_c(A); A\in D,
\end{equation}
where $Y_c(A)$ is the natural parameter for in Weibull response model and $\bm\eta$ is an $n$-dimensional random vector. Similar to \citet{bradley2018bayesian}, we assume the model for count-valued responses to be
\begin{equation}
Z(A_i)\stackrel{ind}\sim Poisson[exp\{Y_d(A_i)\}],
\end{equation}
where the canonical log-link function is used. We make similar assumptions to (\ref{naturalPc}) to incorporate covariate and spatial effects for the Poisson response. That is, the WAP model makes the following assumption:
\begin{equation}
Y_d(A)=\bm x_d(A)'\bm \beta_d+\bm \psi_d(A)'\bm \eta+\bm \psi_d(A)'\bm \eta_d+\gamma_d(A); A\in D,
\end{equation}
where $n_d$-dimensional unknown random vector $\bm \beta_d$ is an unknown $p_d$-dimensional vector of covariate effects, the $r_d$-dimensional $\bm \eta_d$ is assumed to be a MLG random vector (see Appendix B) which is independent with $\bm \eta_c$, and $\bm \eta_d$ captures the independent spatial effect for discrete response, the $r$-dimensional random variable $\bm \eta$ is assumed to be MLG and represents the spatial dependence between Weibull response and Poisson response. 

\begin{figure} 
\caption{A directed graph of the WAP model. The left box represents Weibull responses and the right box represents the Poisson response and its priors. We jointly modeling these two responses with $\bm \eta$ in the middle of the diagram.}
\includegraphics[width=\textwidth]{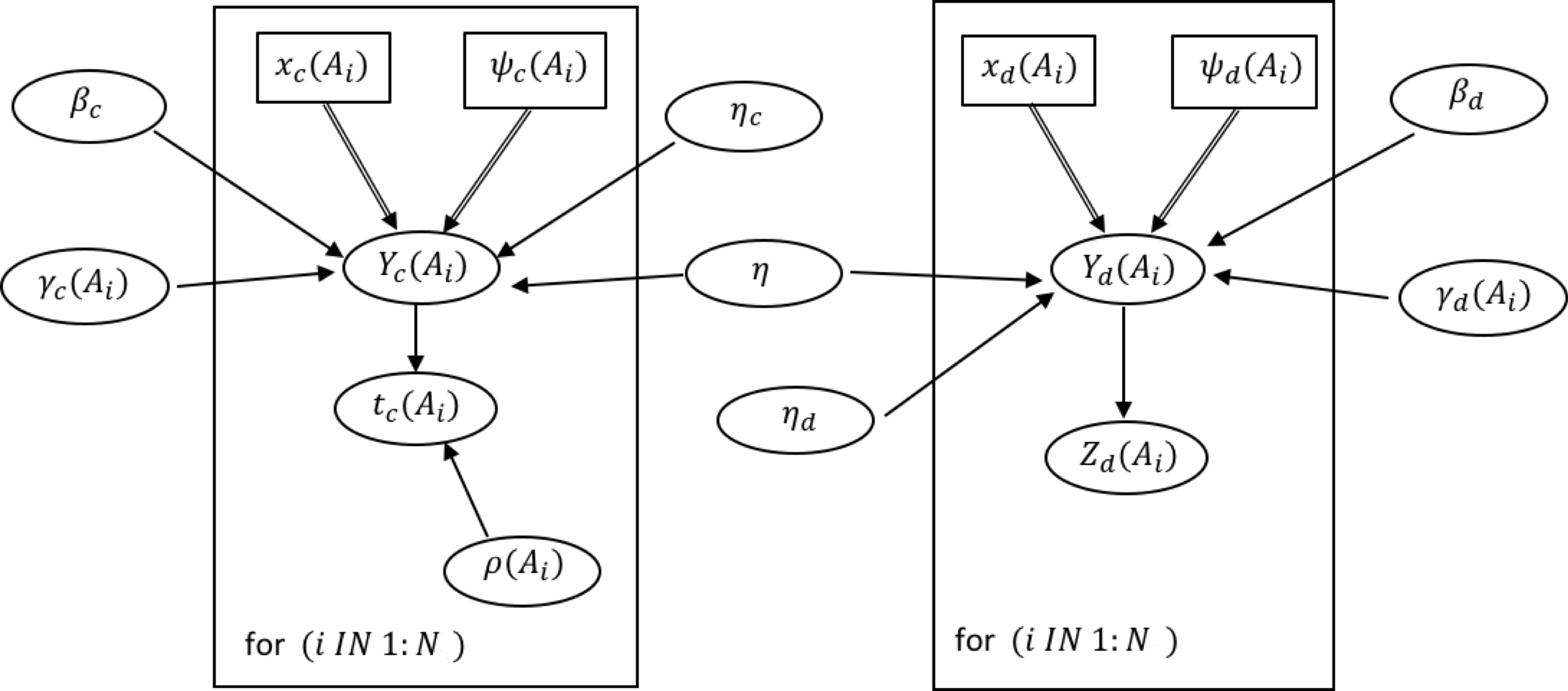}
\label{fig:WAP}
\end{figure}

The $r_d$-dimensional vector $\bm\psi_d(A_i)$ can belong to any class of the areal basis functions, $\gamma_d(A)$ represents the unobserved random effects. The random variable $\bm\eta$ is crucial for obtaining cross-dependence between $Y_c(A_i)$ and $Y_d(A_i)$, which we used as motivation in the introduction. To see this, note that 
\begin{equation*}
Cov\{Y_c(A_i), Y_d(A_i)\}=\bm\psi_c(A_i)'Cov(\bm\eta)\bm\psi_d(A_i),
\end{equation*}
which is not necessarily equal to zero. A complete statement of the WAP model is given in Appendix B. We also provide a directed graph of WAP in Figure~\ref{fig:WAP}. The directed graphs for the univariate response models are the same as Figure~\ref{fig:WAP} with $\bm\eta$ removed.
 We fit this model using a Gibbs sampler. Details can  be found in the Appendix B.

Many of the full-conditional distributions associated with the WAP are conditional MLG distributions, which can be  very difficult to simulate from, and requires iterative algorithms \citep[e.g., slice sampling,][]{neal2003slice} or algorithms with extensive tuning \citep[e.g., Metropolis-Hastings,][]{chib1995understanding}. However, \citet{supbradley2018bayesian} has a data augmentation method that allows one to instead simulate from a marginal distribution of a MLG, which address this issue. We use a similar strategy in this article, for details on this techniques see Appendix D.

\section{Simulation Study \label{simulation study}}
The primary aim of this paper is to jointly analyze PM2.5 (continuous) and mortality (Poisson) to improve the precision of spatial predictions. Thus, in this simulation study, 
we compare to independent (uni-type) analyses of Weibull and Poisson responses in a wide range of scenarios. We also compare the WAP model with a latent Gaussian process model (LGP) to illustrate the advantages of the WAP model. The factors of our simulation study are discussed in Section~\ref{sec:simulation setup}.   

\subsection{Simulation Setup\label{sec:simulation setup}}
In the simulation study, we want to test the predictive performance of the WAP model in several settings. In particular, we track the predictive performance over several choices of the number of basis function, signal to noise ratio (SNR), and the proportion of zeros in the datasets. 
We choose a simulation model that differs from the model we fit the data, to demonstrate the robustness of our parametric assumptions. Specifically, suppose $t(A_i)\sim Weibull(\rho(A_i),exp(-Y_c(A_i))$ and $Z(A_i)\sim Poisson(exp(Y_d(A_i)))$, where
\begin{align}\label{signal}
 &Y_c(A_i)=b_1+c_1 sin(A_i)+\epsilon_1(A_i),\ \epsilon_1(A_i)\stackrel{iid}\sim Normal(0,\sigma^2_c),\nonumber \\
&Y_d(A_i)=b_2+c_2 Y_c(A_i)+\epsilon_2(A_i), \ \epsilon_2(A_i)\stackrel{iid}\sim Normal(0,\sigma^2_d), i=1,\dots,n.
\end{align}
Here, we let $A_i={i}$ and $D=\cup A_i$ where $i=1,\dots,n$. We provide multiple simulations of multi-type spatial fields. We define the SNR to be
\begin{equation*}
SNR_c=\frac{\sum^{n}_{i=1}\{c_1 sin(A_i)\}^2}{n \sigma^2_c},\
SNR_d=\frac{\sum^{n}_{i=1}(c_2 Y_{c,i})^2}{n \sigma_d^2},
\end{equation*}
and, the standard deviation is given by
\begin{equation}
\sigma_c=\sqrt{\frac{\sum^{n}_{i=1}(c_1 sin(A_i))^2}{n SNR_c}}, \
\sigma_d=\sqrt{\frac{\sum^{n}_{i=1}(c_2 Y_{c,i})^2}{n SNR_d}}. 
\end{equation}

For this particular simulation setup when the elements of $\bm\lambda=exp(\bm q_2)$ are consistently less than 0.5, it is highly likely to generate zero counts when simulating from a Poisson with mean $\lambda$. Therefore, we use the following proportion of zeros (POZ) criterion to control for the number of zeros in the data,
\begin{equation*}
POZ=\frac{\# \ of\ elements\ in\ \bm{q}_2\ less\ than\ log(0.5)}{n}.
\end{equation*}
To vary POZ we change the value of $b_2$. We expect better results of WAP model than univariate response models on datasets with small $SNR_c$. This is because a smaller $SNR_c$ leads to smaller variability about the signal. We may expect a similar or even worse performance of WAP model than univariate response models on datasets with large $SNR_c$.

 It is well-known that the number of basis functions can greatly effect the performance of spatial mixed effects models \citep{stein2014limitations}.%%%%%%
Furthermore, the SNR is a well known factor for assessing the predictive performance of functional data \citep{wahba1990spline}. Finally, it is known that an overwhelming number of zeros can lead to difficulties in prediction for spatial statistical models \citep{de2013hierarchical}.
 
In each simulation, we generate 100 Poisson and 100 Weibull responses so that $n=200$ observations.  We define $\bm X$ to be a $n\times p$ covariate matrix with $p=2$, and each element in $\bm X$ is generated from a Bernoulli distribution with success probability equals to 0.5. We do this to mimic the application, which consists of categorical explanation to $Y$. We choose thin plate spline basis functions \citep[i.e.,][$\phi(r)=r^2ln(r)$]{wahba1990spline} to calculate elements in $\bm\Psi_c$ and $\bm\Psi_d$. The true values of shape parameter $\rho$ in Weibull data are generated from Gamma(10,0.1) and we assume 10 adjacent areas shared the same $\rho$. To measure the performance of the predictions, we define the sum of squared error (SSE) as,
\begin{equation*}
SSE_e=(\hat Y_e-Y_e)'(\hat Y_e-Y_e), e=c,d,
\end{equation*}
where $\hat Y_c=\bm {x} ' \hat{\bm {\beta}}_c+\bm\psi_c(\hat{\bm\eta}+\hat{\bm\eta}_c)+\hat\gamma_c$ and $\hat Y_d$ has the same formula. 

\subsection{Sensitivity to Basis Functions \label{sec: simulation basis}}
For illustration, we simulate data according to (\ref{signal})  from signal model with $n=200$, $b_1=-3$, $b_2=8$, $c_1=1.2$, $c_2=1.5$, and both $\sigma_c=\sigma_d=1$. This choice leads to a Poisson dataset with a small number of zeros and a small single noise ratio. We use this setup to simply demonstrate  the ability of WAP model to recover the signals in (\ref{signal}), using different choices of basis functions. In particular, we choose the number of basis function to be 5, 10, and 15. 

\begin{figure}
\caption{These plots shows the predictive performance by choice of basis functions. Panels (a) and (b) show the box plots of total SSE and SSE from $Y_c$ respectively. The white boxes represent the SSE of WAP models and the grey boxes represent the SSE of univariate response models. The x-axis shows the number of basis functions, the y-axis shows the SSE values. Panels (c) and (d) are image examples of results of $Y_c$ and $Y_d$ respectively. Curves are the true values and the red points are estimations (posterior means), and x-axis represents the indices or locations.}
\includegraphics[width=\textwidth]{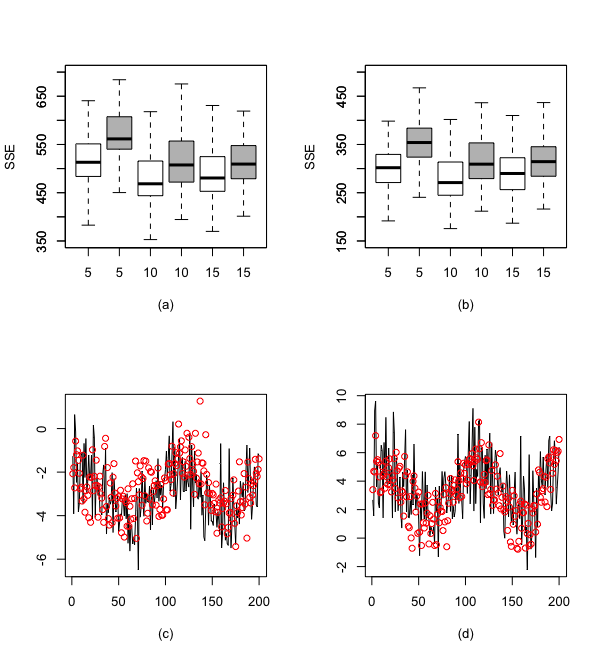}
\label{fig:CompBox}
\end{figure}

In panels (\ref{fig:CompBox}a) and (\ref{fig:CompBox}b), the white boxes present the SSE results for WAP model and the grey boxes display the SSE results of univariate response models. Panel (\ref{fig:CompBox}a) shows that both WAP and univariate response models reach the best performances when using 10 basis functions.  Median SSE for WAP (white boxes) are less or equal to the first quartiles of the SSE of the univariate response model (grey boxes); thus, it appears that WAP is able to recover the signal more so than the univariate response models. Panel (\ref{fig:CompBox}b) shows the comparison of SSE for Weibull models in WAP and univariate response models. It has a similar pattern as the results in Panel (\ref{fig:CompBox}a), again suggesting that WAP performs better than the univariate response Weibull model. Specific examples of the performance of WAP are shown in Panel (\ref{fig:CompBox}c) (Weibull model) and Panel (\ref{fig:CompBox}d) (Poisson model). Here, the circles are the predicted values and curves are true values. We can see that the predicted values generally covers the true values and display the pattern of the unobserved signal.
\begin{table}\label{tab:test1}
\caption{The table shows the p-values from t-tests comparing WAP to uni-type responses on different numbers of basis functions}\label{tab:test1}
  {\begin{tabular}{crrr}
   \hline
   &\multicolumn{3}{c}{p-values} \tabularnewline\
   basis functions & Total SSE & Weibull SSE & Poisson SSE\\
   \hline
   5&6.15$\times 10^{-11}$ & 4.75$\times 10^{-13}$& 0.484 \\
   10& 1.00$\times10^{-3}$ & 6.50$\times 10^{-4}$& 0.525\\
   15& 4.85$\times 10^{-4}$ &1.76$\times 10^{-4}$& 0.534\\
   \hline
  \end{tabular}}
\end{table}
Formal paired t-test are presented in Table~\ref{tab:test1} also provide some evidences. The alternative hypothesis is the that expected SSE when using WAP is smaller than the SSE when using univariate response models. We choose the significance level $0.05$,  and the p-values in Table~\ref{tab:test1} are less than 0.05. Thus, the performance WAP appears to have smaller expected SSE than the univariate response models.

\subsection{Sensitivity to The Signal to Noise Ratio}
Next, we test the performance of WAP in datasets with different SNRs. According to the definition of SNR in Section \ref{sec:simulation setup}, we consider datasets with SNR equal to either 1 or 5. Following Section 3.2, we choose the same specification of (\ref{signal}) and 10 basis functions. 
\begin{figure}
 \caption{These plots shows the predictive performance of different combinations of $SNR_1$ and $SNR_2$.The white boxes represent the SSE of WAP models and the grey boxes represent the SSE of univariate response models. The x-axis in box plots shows the values of signal noise ratio, the y-axis shows the SSE values. (a) shows the box plots of total SSE and (b) illustrates the box plots of SSE from $Y_c$.  (c) and (d) are image examples of results of $Y_c$ and $Y_d$ from $SNR_1=1$ and $SNR_2=5$. Curves are the true values and the circles are estimations (posterior means), and x-axis represents the indices of locations.}
\includegraphics[width=\textwidth]{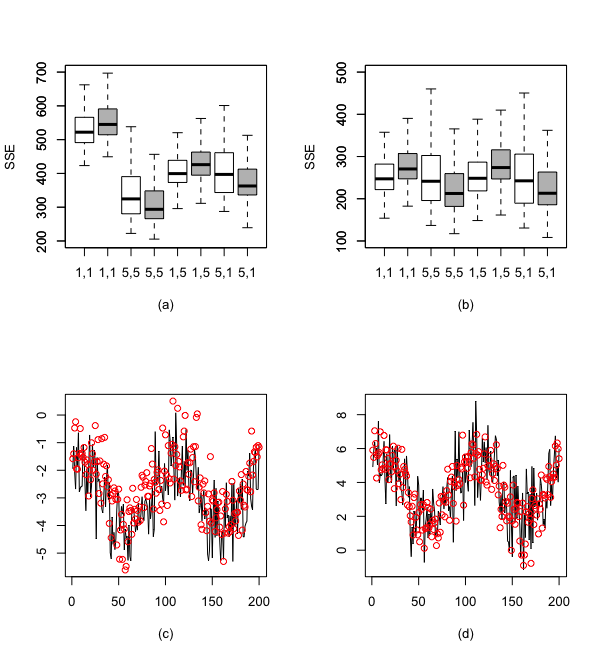}
\label{fig:VarBox}
\end{figure}

Figure (\ref{fig:VarBox}a) shows the SSE associated with WAP (white boxplots) and the SSE associated with univariate response models (grey boxplots) by SNR. When $SNR_c=1$, Panel (\ref{fig:VarBox}a) are the median SSE of WAP (white boxplots) below or equal to the first quartiles of the SSE of the corresponding univariate response model (grey boxplots). Thus, the results of WAP model appears better than the results of univariate response model when $SNR_c=1$. When $SNR_c=5$, WAP and univariate response models appear to perform similarly. Similar patterns are seen in Panel (\ref{fig:VarBox}b), which  presents the SSE for Weibull data by model and SNR.
\begin{table}
\caption{The table shows the p-values from t-tests comparing WAP to uni-type responses on different SNRs}\label{tab:test2}
  {\begin{tabular}{ccrrr}
   \hline
   &&\multicolumn{3}{c}{p-values} \tabularnewline\
   $SNR_c$&$SNR_d$ & Total SSE & Weibull SSE & Poisson SSE\\
   \hline
   1 & 1 & 0.003 & 0.000 & 0.503 \\
   5 & 5 & 0.998 & 0.998 & 0.559 \\
   1 & 5 & 0.012 & 0.013 & 0.509 \\
   5 & 1 & 0.999 & 0.999 & 0.488 \\
   \hline
  \end{tabular}}
\end{table}
Table~\ref{tab:test2} contains the result of the t-tests comparing the SSE between the WAP and univariate response models. At level $0.05$, we can reject the null hypothesis and conclude that the expected SSE is smaller for WAP when $SNR_c=1$. However, we cannot reject the null hypothesis when $SNR_c=5$.
\subsection{Sensitivity to the Proportion of Zero Poisson Counts}
As discussed in the introduction, this type of spatial Poisson models can be sensitive to a large number of zero counts \citep{sellers2016flexible}. In this section, we investigate the performance of our model under different proportion of zero counts in datasets. 

 We keep other specifications fixed and let $SNR_c=1$, $SNR_d=5$. When we generate datasets with $b_2$ equals to 6, 6.5, 7 and 7.5 respectively, the corresponding POZ values tend to be in the respective ranges (11\%,22.5\%), (6\%,17\%), (3\%,11\%), and (1.5\%,7\%). 
\begin{figure} 
\caption{These plots shows the predictive choices by signal zero ratio. The white boxes represent the SSE of WAP models and the grey boxes represent the SSE of univariate response models. The x-axis in box plots shows the values of $b_2$, the y-axis shows the SSE values. Panel (a) shows the box plots of total SSE and panel (b) illustrates the box plots of SSE from $Y_c$.  Panel (c) and panel (d) are image examples of results of $Y_c$ and $Y_d$ from $b_2=6.5$. Curves are the true values and the circles are estimations (posterior mean), and x-axis represents the indices of locations.}
\includegraphics[width=\textwidth]{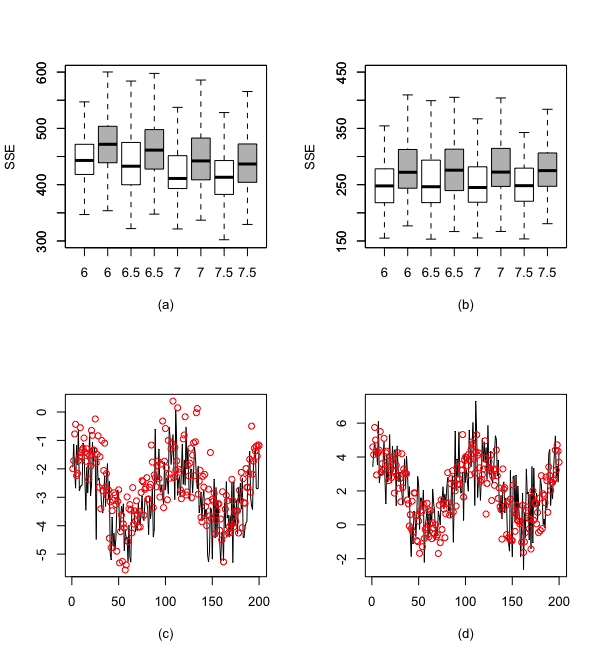}
\label{fig:ZeroBox}
\end{figure}

Figure (\ref{fig:ZeroBox}) provides box plots by model and POZ.  Panel (\ref{fig:ZeroBox}a) and (\ref{fig:ZeroBox}b) display the Poisson SSE and Weibull SSE with the x-axis represents the value of $b_2$. The white box plots present SSE results for WAP  and the grey box plots present SSE results of univariate response models. In Panel (\ref{fig:ZeroBox}a), the (white boxes) median SSE for WAP are very close to the first quartiles of the SSE corresponding to the univariate response models (grey boxes). This shows that the SSE for WAP seems to be smaller than that of the univariate response models. Panel (\ref{fig:ZeroBox}b) shows a similar pattern for Weibull data.
\begin{table}
\caption{The table shows the p-values from t-tests comparing WAP to uni-type responses on different POZs}\label{tab:test3}
  {\begin{tabular}{crrr}
   \hline
   &\multicolumn{3}{c}{p-values} \tabularnewline\
   $b_2$&  Total SSE & Weibull SSE & Poisson SSE\\
   \hline
   6   & 0.001 & 0.000 & 0.501 \\
   6.5 & 0.049 & 0.045 & 0.496 \\
   7   & 0.001 & 0.000 & 0.519 \\
   7.5 & 0.000 & 0.000 & 0.548 \\
   \hline

   $b_2$     & \multicolumn{3}{c}{p-value of WAP comparison}\\
   \hline
   6 and 7.5 & \multicolumn{3}{c}{5.95 $\times10^{-6}$}\\
   \hline
  \end{tabular}}
\end{table}
Table~\ref{tab:test3} contains the t-test results of differences between the WAP SSE and univariate response model SSE. The p-values for Weibull data are all smaller than the significance level $0.05$. We can conclude that WAP appears to do better  than univariate response model on datasets with different POZ values for Weibull data. However, WAP does not have a significant performance when the POZ range is (1.5\%,22.5\%). 

The p-values for Poisson SSEs are not all significant, which confirms to intuition (see discussion in Section 3.1). In summary, the WAP model is preferable to the univariate response model when there is dependence between the datasets and the SNR is not extremely high. Also, the WAP has reasonable performance when the dataset contains approximately 20\% zeros. Since we focus on analyzing federal datasets, where the zero ratio in dataset is very small, this is a reasonable restriction for our purposes. However, a zero-inflated version of WAP is an important topic of future research.

\section{Illustration: Analysis of Mortality and PM2.5 Data from the CDC\label{data analysis} }
The Centers for Disease Control and Prevention provides several types of spatial data including  mortality counts and environmental indicators by US county. In this section, we analyze a big multi-type dataset consisting of air pollution data and mortality counts of diseases related to air quality. Specifically, we analyze averaged Daily Fine Particulate Matter (PM2.5) in $\mu g/m^3$ and mortality counts of diseases related to lung, cardiovascular, respiratory, and stroke in 2011 for all U.S. counties. There is a large literature developing the relationship between PM2.5 and these diseases \citep[e.g.][]{dominici2006fine,franklin2007association,kampa2008human,turner2011long}. WAP explicitly allows researchers to incorporate these types of multi-type dependencies.

For Poisson data, gender (male and female), race (Asian or Pacific Islander, Black or African American and White) and age groups (8 classes from 15 to 85+) are the available covariates. The PM2.5 dataset does not have any immediate  covariate information and hence, is specified so that the intercept is the only covariate. Thus $p_d=13$ and $p_c=1$.  The CDC provides 12,760 mortality counts and 3,111 PM2.5 values. We use the deviance information criterion (DIC; \citet{spiegelhalter2002bayesian}) to choose the number of basis functions. We consider the number of basis functions to be 20, 30, 40 and 50, and the value with smallest DIC is $r=40$. We assume $\delta_i$ to be constant across each state, and the remaining prior settings are kept the same as they were in our simulation study. We also use the bisquare basis function from from \citet[][]{cressie2008fixed}. The bisquare basis function is defined as,
\begin{align}
\psi_j(A_i)\equiv \Big\{1-\Big (\frac{||{\bm{u}_i}-{\bm{v}_j}||}{r_l}\Big)^2\Big\}^2, \ for \ ||{\bm{u}_i}-{\bm{v}_j}|| < r_l;i=1,\dots,N, j=1,\dots,r.
\end{align}
where $\bm{u}_i$ is the centroid of county $A_i$, $\bm{v}_j$ is $i$th knot location and $r_l$ is 1.5 times the median of distance between knots in the set $\{\bm u_i: i=1,\dots, r\}$. We find that using the median between knot distances to define $r_l$ can produce a smaller DIC for this dataset than the minimum between knot distances, as used in \citet[][]{cressie2008fixed}. We run the MCMC algorithm for 30,000 iterations and burn-in the first 20,000 iterations. The convergence is verified by trace plots.

\begin{figure} \label{fig:WAPQQ}
\caption{This figure shows scatterplots and histograms of residual from the WAP model on a data analysis with 40 basis functions. Panel (a) shows the scatterplots of PM2.5 and Panel (b) displays the scatterplots based on the log of predictions and log observations of deaths. We have added $log\big\{\frac{E(R)}{E(\sqrt R)^2}\big\}$ from \citet{aldaz2008selfimprovemvent}, where $R$ is a posterior replicate predictive of the response. Panel (c) and (d) are the histograms of the residuals for Weibull and Poisson responses from WAP model.}
\includegraphics[width=\textwidth]{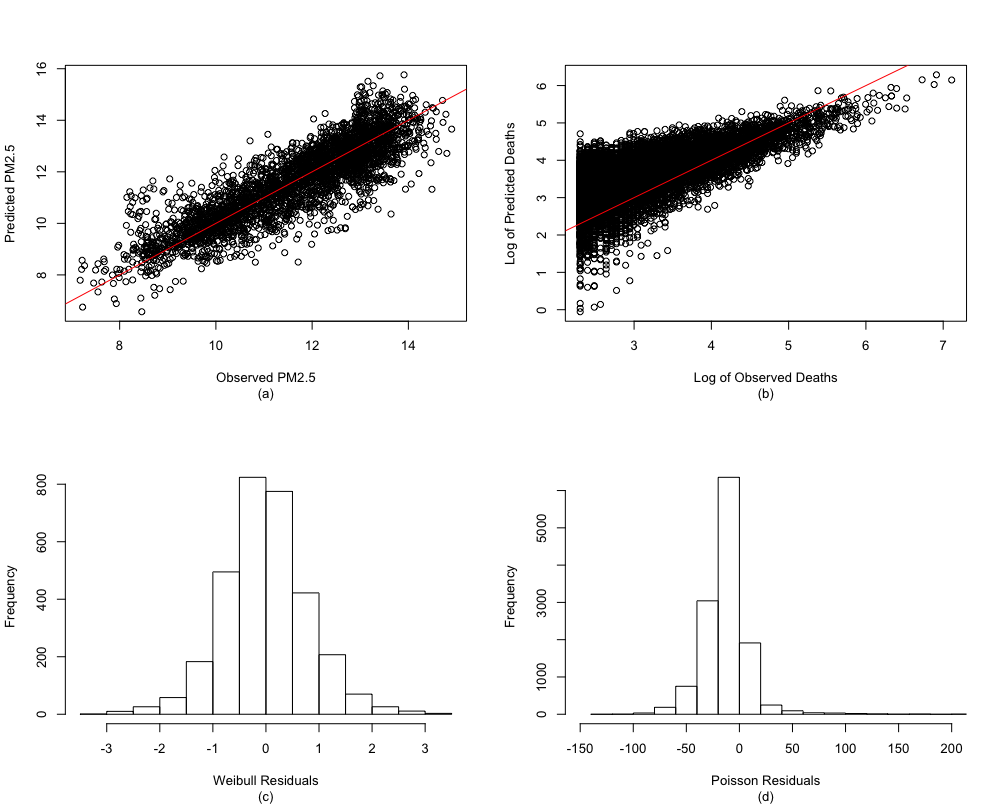}
\end{figure}

In our data analysis, we calculate the quantiles of elements in $\bm\eta$ to investigate the need for multi-type dependence. We find  that of 14 elements have point wise credible intervals that are greater than zero, and 7 elements have point wise credible intervals that are completely less than zero. This shows that the data provide evidence for incorporating cross-response type dependence. We tried different types of basis functions and all the results support multi-response dependence. The mean squared errors (MSE) based on the expectations and observations of Weibull and Poisson (with log transform)  model are 0.493 and 0.182 respectively. Here MSE is the mean squared error between the observations and the posterior predicted values. Figure~\ref{fig:WAPQQ} shows both scatterplots of the estimated values (posterior means) versus the observations (Poisson on log scale), and histograms of the residuals by responses type. \citet{aldaz2008selfimprovemvent} includes a correction term when interpreting the expected value of the data on the log-scale, which we include in the Panels (b) in Figure~\ref{fig:WAPQQ}. The scatterplots for both response types suggest  that the WAP model performs well and the residual plots look slightly symmetric and unimodel. The Poisson residuals have a slight right skewness. This suggests that we have smoothed our Poisson estimates, which motivates us to consider the posterior predictive p-value to determine if we have over-smoothed our Poisson estimates.

To assess the goodness-of-fit of WAP model, we use the posterior predictive p-value \citep{meng1994posterior,gelman1996posterior} with the Chi-squared distribution. Here the posterior predictive p-value is computed as: 
\begin{equation}\label{chi-stata}
\frac{1}{B}\sum^{B}_{b=1} I(\chi^2_b>\chi^2_o)
\end{equation}
where $B=10000$, $I(\cdot)$ is the indicator function , $\chi^2_b$ is the Chi-statistic between the posterior mean of the responses and the $b$-th replicate of the posterior predictive distribution, and $\chi^2_o$ is the Chi-square statistic between the posterior mean of the responses and the observations.

The posterior predictive p-value of the Weibull responses in WAP model and Univariate Weibull model are 0.670 and 0.989 respectively. The posterior predictive p-value of the Poisson responses in WAP model and Univariate Poisson models are 0.568 and 0.510 respectively. These posterior predictive p-values are computed based on (\ref{chi-stata}). These values suggest that the WAP model slightly overfits the data but still has strong out-of-sample performance. Moreover, the uni-type Weibull model overfit the data and the uni-type Poisson data slightly overfit the data.

We present the predictions for the state of California for the visualization purposes. Figure~\ref{fig:mapsPM} shows  maps of raw data and predictions from WAP for PM2.5 over counties in California. We see that both maps show an increasing trend of PM2.5 from the west to east. As expected, some counties have slightly higher predicted  values than the raw data value. This is because we are estimating the mean of the Weibull, while the data are assumed to be realizations from a Weibull.
\begin{figure} \label{fig:mapsPM}
\caption{Panel (a) shows the map of raw PM2.5 data and Panel (b) illustrates the posterior mean of PM2.5 over counties in California. }
\includegraphics[width=\textwidth]{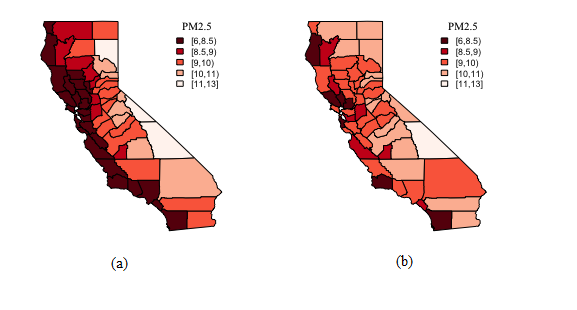}
\label{fig:mapsPM}
\end{figure}

\begin{figure}\label{fig:mapsMort}
\caption{Panel (a) shows the map with raw hazard and Panel (b) illustrates the posterior mean hazard for white women over 85 who suffered from lung cancer in California. }
\includegraphics[width=\textwidth]{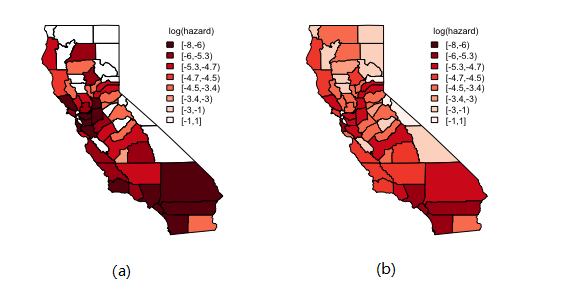}
\label{fig:mapsPM}
\end{figure}

We present our results for white females over 85 years old who suffered from the lung cancer. This particular group had the most number of observations. Figure~\ref{fig:mapsMort} presents the county-level maps of real data and the predicted values for category. In Panel (\ref{fig:mapsMort}a), we plot the hazard of the raw data. The hazard is computed as follows:
\begin{equation}
hazard=\frac{mortality}{population}.
\end{equation}
We again see an increasing trend, where fewer deaths occur on the west coast than on the east areas.

By comparing the predicted maps of PM2.5 and the predicted hazard rate of white females over 85 years old, we see that the   hazard rate is increasing from the west to east in general, which is similar to the trend of PM2.5. This adds additional evidence to the lung disease literature that PM2.5 is related to lung diseases.

\section{Discussion\label{discussion}}
In this article we are motivated by CDC dataset. In particular, monitoring PM2.5 is important because it helps to assess public health and provides an avenue to do the clinical inference related to mortality. These variables are known to be dependent \citep{laden2000association,schwartz2000fine,valavanidis2008airborne}. As a result we developed the WAP to leverage multi-type dependence to improve estimation of PM2.5.
We introduce the joint Weibull and Poisson (WAP) model, which is a framework that can be used to model high-dimensional continuous and count-valued (or multi-type) responses. Most of the work  modeling multi-type responses are in machine learning and nonparametric settings, and hence, our WAP fills a gap within the Bayesian analysis literature. Another important contribution is that we use the multivariate log-gamma distribution as a conjugate prior. This allows the WAP to be easily implemented using a collapsed Gibbs sampler, which avoids complicated tuning and other  approaches used in other standard Bayesian algorithms. Using a reduced rank set of basis functions, our WAP can be applied to high-dimensional datasets with less cost than full-rank methods.

In the simulations study, we test the sensitivities of WAP to several different factors including the choice of the number of basis functions, the signal to noise ratio, and the proportion of zero Poisson counts. We generate data that is different from our model, and  the WAP was able to accurately estimate this signal even though the data were not generate from WAP.  After comparing to univariate response models, the WAP appeared to have much stronger predictive performance in  most situations.  

The simulation study suggested that the performance of WAP may decrease when faced with a dataset with a large proportion of zeros. This result is expected since there is a rather large literature on zero-inflated Poisson models, which are motivated by similar empirical results. Consequently a zero-inflated WAP is an important topic of future research.
\vfill

\appendix
\section{A: Uni-Type Model for Weibull Responses}
The statistical model for a univariate response Weibull data is the product of, 
\begin{align}\label{weibullModel}
&Data\ Model: t(A_i)|\bm{\beta,\eta},\gamma(A_i) \stackrel{ind}\sim Weibull\big(\rho(A_i,\bm\delta),exp[-\{\bm x (A_i)'\bm \beta+\bm{\psi}(A_i)'\bm\eta+\gamma(A_i)\}]\big)\nonumber\\
&Process\ Model: \bm \eta \sim MLG\big(\bm{0},\bm{V}, \alpha_{\eta}\bm{1}_n,\kappa_{\eta}\bm{1}_n\big)\nonumber\\
&Parameter\ Model\ 1: \bm\beta \sim MLG\big(\bm{0}, \bm{I_p}, \alpha_{\beta}\bm{1}_p,\kappa_{\beta}\bm{1}_p\big)\nonumber\\
&Parameter \ Model \ 2: \bm\gamma \sim MLG\big(\bm{0}, \bm{I}_n, \alpha_{\beta}\bm{1}_n,\kappa_{\beta}\bm{1}_n\big)\nonumber\\
&Parameter \ Model \ 3: V_{sj}\sim LG(\alpha_v,\kappa_v), \ s=1,\dots, r; \ j=1,\dots, r;\nonumber\\ 
&Parameter \ Model \ 4: (\alpha_{\eta},\kappa_{\eta})\sim f(r_1,r_2,b)\nonumber\\ 
&Parameter \ Model \ 5: (\alpha_{\beta},\kappa_{\beta})\sim f(r_1,r_2,b)\nonumber\\ 
&Parameter \ Model \ 6: (\alpha_{\gamma},\kappa_{\gamma})\sim f(r_1,r_2,b),\nonumber\\ 
&Parameter \ Model \ 7: \bm\delta \stackrel{ind} \sim Gamma(\alpha_{\rho},\kappa_{\rho}), \ i=1,\dots, n, 
\end{align}
In (\ref{weibullModel}) we list the conditional distributions and marginal distributions, whose product gives a joint distribution, which we use for inference.
In (\ref{weibullModel}) $f(r_1,r_2,b)$ stand for the conjugate prior of shape and scale parameters developed in \citet{bradley2018bayesian}. Specifically, 
 \begin{equation*}
 f(r_1,r_2,b)=\frac{1}{\Gamma(\alpha)^b}(\kappa^{-b}exp(r_1))^{\alpha}exp(r_2\kappa)
 \end{equation*}
where $r_1>0$, $r_2<0$. We assume that $\bm{V}^{-1}$ is a lower-diagonal modified Cholesky matrix with unit diagonal and $(s,j)-th$ element denoted with $V_{sj}$ for $s>j$.
  to form a conjugate posterior distribution for each pair of $\alpha$ and $\kappa$.
The shape parameter $\bm{\rho}(\bm\delta)=(\rho(A_1,\bm\delta),\dots,\rho(A_n,\bm\delta))'$ is modeled using a basis function expansion,
\begin{equation*}
\bm{\rho}(\bm\delta)=\bm{\Phi}\bm \delta,
\end{equation*}
where the $n\times m$ matrix $\bm\Phi$ consists of zeros and ones, and $m\ll n$. In each row there is only a single one that is present. This essentially defines a region specific shape parameter. In our simulation we treat $\bm\Psi$ as known, and let $\bm\Psi$ consist of state-level indicators in our application. In Section 4, we assume each U.S. state has the same shape parameter. 

The WAP model is then defined to be proportional to the product of the following conditional and marginal distributions:
\begin{align}\label{wap}
&Data \ Model\ 1: t(A_i)|\bm{\beta}_c,\bm{\eta}_c,\bm\eta,\gamma_c(A_i) \stackrel{ind}\sim Weibull\big(\rho(A_i,\bm\delta),exp\{-Y_c(A_i)\}\big)\nonumber\\
&\qquad \qquad \qquad \qquad Y_c(A_i)=\bm x(A_i)'\bm \beta+\bm{\psi}(A_i)'(\bm\eta+\bm\eta_c)+\gamma(A_i)\nonumber\\
&Data \ Model\ 2: Z(A_i)|\bm{\beta}_d,\bm\eta_d,\bm\eta,\gamma_d(A_i) \stackrel{ind}\sim Poisson\big(exp\{Y_d(A_i)\}\big)\nonumber\\
&\qquad \qquad \qquad \qquad Y_d(A_i)=\bm x(A_i)'\bm \beta_d+\bm{\psi_d}(A_i)'(\bm\eta+\bm\eta_d)+ \gamma_d(A_i)\nonumber\\
&Process\ Model \ 1: \bm \eta \sim cMLG\big(\bm 0, \bm{V}_{\eta}, \bm \alpha_{\eta}^*,\bm\kappa_{\eta}^*\big), \bm \alpha_{\eta}^*=(\zeta\bm{1}_r',\alpha_{\eta}\bm{1}_n')', \bm \kappa_{\eta}^*=(\zeta\bm{1}_r',\kappa_{\eta}\bm{1}_n')'\nonumber\\
&Process\ Model \ 2: \bm \eta_c \sim MLG\big(\bm{0},\bm V, \alpha_{\eta_c}\bm{1}_n,\kappa_{\eta_c}\bm{1}_n\big)\nonumber\\
&Process\ Model \ 3: \bm \eta_d \sim cMLG\big(\bm 0, \bm{V}_{\eta_d}, \bm\alpha_{\eta_d}^*,\bm\kappa_{\eta_d}^*\big), \bm \alpha_{\eta_d}^*=(\zeta\bm{1}'_r,\alpha_{\eta_d}\bm{1}'_n)', \bm \kappa_{\eta_d}^*=(\zeta\bm{1}_r',\kappa_{\eta_d}\bm{1}_n')'\nonumber\\
&Parameter\ Model\ 1: \bm\beta_c \sim MLG\big(\bm{0}, \bm I_p, \alpha_{\beta_c}\bm{1}_p,\kappa_{\beta_c}\bm{1}_p\big)\nonumber\\
&Parameter\ Model\ 2: \bm\beta_d \sim cMLG\big(\bm 0, \bm{V}_{\beta_d}, \bm\alpha^*_{\beta_d},\bm\kappa^*_{\beta_d}\big), \bm \alpha_{\beta_d}^*=(\zeta\bm{1}_r',\alpha_{\beta_d}\bm{1}_n')', \bm \kappa_{\beta_d}^*=(\zeta\bm{1}_r',\kappa_{\beta_d}\bm{1_n}')'\nonumber\\
&Parameter \ Model \ 3: \bm\gamma_c \sim MLG\big(\bm{0},\bm I_n, \alpha_{\gamma_c}\bm{1}_n,\kappa_{\gamma_c}\bm{1}_n\big)\nonumber\\
&Parameter \ Model \ 4: \bm\gamma_d \sim cMLG\big(\bm 0, \bm{V}_{\gamma_d}, \bm\alpha_{\gamma_d}^*,\bm\kappa^*_{\gamma_d}\big), \bm \alpha_{\gamma_d}^*=(\zeta\bm{1}_r',\alpha_{\gamma_d}\bm{1}_n')', \bm \kappa_{\gamma_d}^*=(\zeta\bm{1}_r',\kappa_{\gamma_d}\bm{1}_n')'\nonumber\\
&Parameter \ Model \ 5: V_{sj}\sim LG(\alpha_v,\kappa_v), \ s=1,\dots, r; \ j=1,\dots, r;\nonumber\\ 
&Parameter \ Model \ 6: V_{d(sj)}\sim LG(\alpha_v,\kappa_v), \ s=1,\dots, r; \ j=1,\dots, r;\nonumber\\ 
&Parameter \ Model \ 7: V_{c(sj)}\sim LG(\alpha_v,\kappa_v), \ s=1,\dots, r; \ j=1,\dots, r;\nonumber\\ 
&Parameter \ Model \ 8: (\alpha_{a},\kappa_{a})\sim f(r_1,r_2,b),\ a=\eta, \eta_c,\eta_d, \beta_c,\beta_d,\gamma_c,\gamma_d\nonumber\\ 
&Parameter \ Model \ 9: \bm\delta \stackrel{ind}\sim Gamma(\alpha_{\rho},\kappa_{\rho}), \ i=1,\dots, n,
\end{align}
where $\bm V_{\eta}= (\bm \Psi_d',\bm V')'$ in Process Model 1, $\bm{V}_{\eta_d}=(\bm X',\bm V_d')'$ in Process Model 3, $\bm{V}_{\beta_d}=(\bm X',\bm I'_p)'$ in Parameter Model 2, $\bm{V}_{\gamma_d}=(\bm I_n', \bm I'_n)'$ in Parameter Model 4, and $\zeta$ is a very small constant to avoid boundary values.  The model in Appendix B is proportional to (\ref{wap}) when conditioning on all $\bm q_{sub}$ values equal to zero. This is why we condition on zero for the augmented values in the Gibbs sampler.
We give cMLG prior to those parameters that are correlated with discrete data in order to avoid these computational issues with tuning or rejection based algorithms.
We also model the shape parameter of the Weibull distribution. In many settings it is assumed that the shape parameter is known \citep[i.e.,][]{nassar2005bayesian}, and thus, our model offers a straight forward approach to estimate this parameter.

\section{B: WAP Model Prior Specifications and the Derivation of the Full-Conditional Distributions}
The WAP model with data augmentation is as follow:
\begin{align*}
&Data \ Model\ 1: t(A_i)|\bm{\beta}_c,\bm{\eta}_c,\bm\eta,\gamma_c(A_i) \stackrel{ind}\sim Weibull\big(\rho(A_i,\bm\delta),exp\{-Y_c(A_i)\}\big)\nonumber\\
&\qquad \qquad \qquad \qquad Y_c(A_i)=\bm x_c(A_i)'\bm \beta_c+\bm{\psi}_c(A_i)'(\bm\eta+\bm\eta_c)+\gamma_c(A_i)+\bm Q_{1\eta}(A_i)'\bm q_{\eta}+\dots+Q_{1v_{\eta_c(r)}}q_{v_{\eta_c(r)}}\nonumber\\
&Data \ Model\ 2: Z(A_i)|\bm{\beta}_d,\bm\eta_d,\bm\eta,\gamma_d(A_i) \stackrel{ind}\sim Poisson\big(exp\{Y_d(A_i)\}\big)\nonumber\\
&\qquad \qquad \qquad \qquad Y_d(A_i)=\bm x_d(A_i)'\bm \beta_d+\bm{\psi_d}(A_i)'(\bm\eta+\bm\eta_d)+ \gamma_d(A_i)+\bm Q_{2\eta}(A_i)'\bm q_{\eta}+\dots+Q_{2v_{\eta_d(r)}}q_{v_{\eta_d(r)}}\nonumber\\%%
&Process\ Model \ 1: \bm \eta \sim cMLG\big(\bm c_{\eta}, \bm{H}^*_{\eta}, \bm \alpha_{\eta}^*,\bm\kappa_{\eta}^*\big), \bm \alpha_{\eta}^*=(\bm\zeta',\bm\alpha'_{\eta})', \bm \kappa_{\eta}^*=(\bm\zeta',\bm\kappa'_{\eta})'\nonumber\\
&\qquad \qquad \qquad \qquad \bm c_{\eta}=-\bm V^{*^{-1}}_{\eta}\bm Q^*_{\eta}\bm q_{\eta}-\bm V^{*^{-1}}_{\eta}(\bm Q^*_{1v_{\eta}}\bm q^*_{v_{\eta}}), \bm Q^*_{\eta}=[\bm Q'_{2\eta},\bm Q'_{3\eta}]'\nonumber\\
& \qquad \qquad \qquad \qquad\bm Q^*_{1v_{\eta}}=(\bm 0_{},\bm Q_{1v_{\eta(j)}}), \bm q^*_{v_{\eta}}=(\bm 0', \bm q_{v_{\eta(j)}}')'\\
&Process\ Model \ 2: \bm \eta_c \sim MLG\big(\bm c_{\eta_c},\bm H^*_{\eta_c}, \bm\alpha'_{\eta_c},\bm\kappa'_{\eta_c}\big), \bm c_{\eta_c}=-\bm V^{-1}_{\eta_c}\bm Q_{2\eta_c}\bm q_{\eta_c}-\bm V^{*^{-1}}_{\eta_c}(\bm Q^*_{1v_{\eta_c}}\bm q^*_{v_{\eta_c}})\nonumber\\
& \qquad \qquad \qquad \qquad\bm Q^*_{1v_{\eta_c}}=(\bm 0,\bm Q_{1v_{\eta_c(r)}}), \bm q^*_{v_{\eta_c}}=(\bm 0', \bm q_{v_{\eta_c(r)}}')'\\
&Process\ Model \ 3: \bm \eta_d \sim cMLG\big(\bm c_{\eta_d}, \bm{H}^{*}_{\eta_d}, \bm\alpha_{\eta_d}^*,\bm\kappa_{\eta_d}^*\big), \bm \alpha_{\eta_d}^*=(\bm\zeta',\bm\alpha'_{\eta_d})', \bm \kappa_{\eta_d}^*=(\bm\zeta',\bm\kappa_{\eta_d}')'\nonumber\\
& \qquad \qquad \qquad \qquad \bm c_{\eta_d}=-\bm V^{*^{-1}}_{\eta_d}\bm Q_{\eta_d}\bm q_{\eta_d}-\bm V^{*^{-1}}_{\eta_d}(\bm Q^*_{1v_{\eta_d}}\bm q^*_{v_{\eta_d}})\\
& \qquad \qquad \qquad \qquad\bm Q^*_{1v_{\eta_d}}=(\bm 0,\bm Q_{1v_{\eta_d(j)}}), \bm q^*_{v_{\eta_d}}=(\bm 0', \bm q_{v_{\eta_d(j)}}')'\\
&Parameter\ Model\ 1: \bm\beta_c \sim MLG\big(\bm c_{\beta_c}, \bm I_{\beta_c}, \bm\alpha_{\beta_c},\bm\kappa_{\beta_c}\big), \bm c_{\beta_c}=-\bm Q_{2\beta_c}\bm q_{\beta_c}\nonumber\\
&Parameter\ Model\ 2: \bm\beta_d \sim cMLG\big(\bm c_{\beta_d}, \bm{H}^{*}_{\beta_d}, \bm\alpha^*_{\beta_d},\bm\kappa^*_{\beta_d}\big), \bm \alpha_{\beta_d}^*=(\bm\zeta',\bm\alpha'_{\beta_d})', \bm \kappa_{\beta_d}^*=(\bm\zeta',\bm\kappa'_{\beta_d})'\nonumber\\
& \qquad \qquad \qquad \qquad \bm c_{\beta_d}=-\bm V^{*^{-1}}_{\beta_d}\bm Q_{\beta_d}\bm q_{\beta_d}\\
&Parameter \ Model \ 3: \bm\gamma_c \sim MLG\big(\bm c_{\gamma_c},\bm I_n, \bm\alpha'_{\gamma_c},\bm\kappa_{\gamma_c}'\big), \bm c_{\gamma_c}=-\bm Q_{2\gamma_c}\bm q_{\gamma_c}\nonumber\\
&Parameter \ Model \ 4: \bm\gamma_d \sim cMLG\big(\bm c_{\gamma_d}, \bm{H}^{*}_{\gamma_d}, \bm\alpha_{\gamma_d}^*,\bm\kappa^*_{\gamma_d}\big), \bm \alpha_{\gamma_d}^*=(\bm\zeta',\bm\alpha_{\gamma_d}')', \bm \kappa_{\gamma_d}^*=(\bm\zeta',\bm\kappa_{\gamma_d}')'\nonumber\\
& \qquad \qquad\qquad\qquad \bm c_{\gamma_d}=-\bm V^{*^{-1}}_{\gamma_d}\bm Q_{\gamma_d}\bm q_{\gamma_d}\\
&Parameter \ Model \ 5: V_{\eta(sj)}\sim MLG(c_{\eta(j)},1,\alpha_v,\kappa_v), \ s=1,\dots, r; \ j=1,\dots, r;\nonumber\\ 
&Parameter \ Model \ 6: V_{\eta_d(sj)}\sim MLG(c_{\eta_c(j)},1,\alpha_v,\kappa_v), \ s=1,\dots, r; \ j=1,\dots, r;\nonumber\\ 
\end{align*}

\begin{align*}
&Parameter \ Model \ 7: V_{\eta_c(sj)}\sim MLG(c_{\eta_d(j)},1,\alpha_v,\kappa_v), \ s=2,\dots, r; \ j=1,\dots, r-1;\qquad \qquad \qquad\qquad\nonumber\\ 
&Parameter \ Model \ 8: (\alpha_{a},\kappa_{a})\sim f(r_1,r_2,b),\ a=\eta, \eta_c,\eta_d, \beta_c,\beta_d,\gamma_c,\gamma_d\nonumber\\ 
&Parameter \ Model \ 9: \bm\delta \stackrel{ind}\sim Gamma(\alpha_{\rho},\kappa_{\rho}), \ i=1,\dots, n,
\end{align*}
where $\bm Q_{\eta}\bm q_{\eta}+\dots+Q_{1v_{\eta_c(r)}}q_{v_{\eta_c(r)}}$ is the sum over all augmented values (e.g., see Appendix D) $\bm q_{\eta}, \bm q_{\eta_c}, \dots, \bm q_{\eta_c}(r)$, which are assumed to have improper prior 1 (e.g., $f(\bm q_{\eta})=1$).

Let $\bm V^{*^{-1}}_{\eta}= [\bm H^*_{\eta}, \bm Q_{\eta}]$, $\bm H^*_{\eta}=[\bm \Psi_d',\bm V_{\eta}']'$, $\bm{V}^{*^{-1}}_{\eta_d}=[\bm H^*_{\eta_d}, \bm Q_{\eta_d}]$, $\bm H^*_{\eta_d}=[\bm\Psi_d',\bm V_d']'$, $\bm{V}^*_{\beta_d}=[\bm H^*_{\beta_d},\bm Q_{\beta_d}]$, $\bm H^*_{\eta_d}=[\bm X_d',\bm I'_{\beta_d}]'$, $\bm{V}_{\gamma_d}=[\bm I_n', \bm I'_n]'$, $c_{\eta(j)}=-\bm V^{*^{-1}}_{\eta}Q_{1\eta(j)}q_{\eta(j)}$, $c_{\eta_c(j)}=-\bm V^{*^{-1}}_{\eta_c}Q_{1\eta_c(j)}q_{\eta_c(j)}$, $c_{\eta_d(j)}=-\bm V^{*^{-1}}_{\eta_d}Q_{1\eta_d(j)}q_{\eta_d(j)}$, and $\zeta$ is a very small constant to have well-defined full-conditional distributions when count-values are zeros.
In general, $\bm Q_{sub}$ is the basis of the null space associated with $\bm H^*_{sub}$, where $sub=\eta$, $\eta_c$, $\eta_d$, and $\beta_d$. In this case, the marginal distribution of an MLG with precision parameter $V_{sub}$ is given by $(\bm H^{*'}_{sub}\bm H^{*}_{sub})^{-1}\bm H^{*'}_{sub}\bm w$, where $\bm w$ is MLG with identity precision parameter (see Appendix D for an example). 

We use the full-conditional distribution of $\bm \eta$ as an example here:
\begin{align*}
&p(\bm \eta,\bm q_{\eta}|\bm t, \bm Z, \bm V_{\eta},\bm\alpha_{\eta}, \bm\kappa_{\eta},\bm q_{\eta_c}=\bm 0,\dots,q_{v_{d(sj)}}=0)\\
\propto & f(\bm t|\bm Z, \bm{\beta}_c,\bm{\eta}_c,\bm\eta,\bm\gamma_c,\bm q_{\eta_c}=\bm 0,\dots,q_{v_{\eta_c(r)}}=0)f(\bm Z|\bm{\beta}_d,\bm\eta_d,\bm\eta,\bm\gamma_d,\bm q_{\eta_d}=\bm 0,\dots,q_{v_{\eta_d(r)}}=0)\\
&f(\bm \eta|\bm q_{\eta},\bm V_{\eta},\bm\alpha_{\eta},\bm\kappa_{\eta})f(\bm q_{\eta})\\
\propto & exp\big\{\bm 1'_{\eta}\bm\Psi_c\bm\eta+\bm Z'\bm\Psi_d\bm\eta+\bm\alpha_{\eta}^{*'}\bm V^*_{\eta}\bm\eta+\bm 1'_{\eta}\bm Q_{1\eta}\bm q_{\eta}+\bm Z'\bm Q_{2\eta}\bm q_{\eta}-\bm\alpha^{*'}_{\eta}\bm V^*_{\eta}\bm c_{\eta}-\bm\kappa_{\eta}^{*'}exp(\bm V^*_{\eta}\bm\eta-\bm V^*_{\eta}\bm c_{\eta})\big\}\\
& -(\bm t^{\bm\rho})'exp(\bm X_c\bm\beta_c+\bm\Psi_c\bm\eta+\bm\Psi_c\bm\eta_c+\bm\gamma_c+\bm Q_{1\eta}\bm q_{\eta})-\bm 1'_{\eta} exp(\bm X_d\bm\beta_d+\bm\Psi_d\bm\eta_d+\bm\Psi_d\bm\eta+\bm\gamma_d+\bm Q_{2\eta}\bm q_{\eta})\\
\propto & exp\big\{\bm 1'_{\eta}\bm\Psi_c\bm\eta+(\bm Z'+\bm\zeta')\bm\Psi_d\bm\eta+\bm\alpha'_{\eta}\bm V_{\eta}\bm\eta+\bm 1'_{\eta}\bm Q_{1\eta}\bm q_{\eta}+(\bm Z'+\bm\zeta')\bm Q_{2\eta}\bm q_{\eta}+\bm\alpha'_{\eta}\bm Q_{3\eta}\bm q_{\eta}\\
&-(\bm t^{\bm \rho}\circ exp(\bm X_c\bm\beta_c+\bm\Psi_c\bm\eta_c+\bm\gamma_c))'exp(\bm\Psi_c\bm\eta+\bm Q_{1\eta}\bm q_{\eta})\\
& -(exp(\bm X_d\bm\beta_d+\bm\Psi_d\bm\eta_d+\bm\gamma_d)'+\bm\zeta')exp(\bm\Psi_d\bm\eta+\bm Q_{2\eta}\bm q_{\eta})-\bm\kappa'_{\eta}exp(\bm V_{\eta}\bm\eta+\bm Q_{3\eta}\bm q_{\eta})\big\}\\
\propto & exp\big\{\bar{\bm\alpha}'_{\eta}[\bm H_{\eta},\bm Q_{\eta}] (\bm\eta',\bm q'_{\eta})'-\bar{\bm\kappa}'_{\eta}exp([\bm H_{\eta},\bm Q_{\eta}](\bm\eta',\bm q'_{\eta})')\big\}\\
\propto & MLG(\bm 0,[\bm H_{\eta},\bm Q_{\eta}]^{-1},\bar{\bm\alpha}_{\eta},\bar{\bm\kappa}_{\eta})
\end{align*}
where $\bm H_{\eta}=[\bm\Psi'_c,\bm\Psi'_d,\bm V_{\eta}']'$, $\bm Q_{\eta}=[\bm Q'_{1\eta},\bm Q'_{2\eta}, \bm Q'_{3\eta}]'$, $\bar{\bm\alpha}_{\eta}=(\bm 1'_{\eta},\bm Z'+\bm\zeta',\bm\alpha'_{\eta})'$, $\bar{\bm\kappa}=((\bm t^{\bm\rho}\circ exp(\bm X_c\bm\beta_c+\bm\Psi\bm\eta_c+\bm\gamma_c))',exp(\bm X_d\bm\beta_d+\bm\Psi_d\bm\eta_d+\bm\gamma_d)'+\bm\zeta', \bm\kappa'_{\eta})'$.
To simulate from $f(\bm \eta |\bm t,\bm Z, \bm V_{\eta}, \bm \alpha_{\eta}, \bm \kappa_{\eta}, \bm q_{\eta}=\bm 0,\dots, q_{v_d(r)}=0)$ on can compute $(\bm H'_{\eta} \bm H_{\eta})^{-1}\bm H'_{\eta}\bm w$ where $\bm w\sim MLG(\bm 0, \bm I_{\eta}, \bar{\bm\alpha}_{\eta},\bar{\bm\kappa}_{\eta})$.

Similar to the procedure of achieving the posterior of $\bm \eta$, we can have posteriors of other parameters. 

2.The full-conditional distribution of $\bm\eta_c$ is:

$p(\bm\eta_c,\bm q_{\eta_c}|\bm t,\bm\eta,\bm\beta_c,\bm\gamma_c,\bm V_{\eta_c},\bm\alpha_{\eta_{c}},\bm\kappa_{\eta_{c}},\bm q_{\eta}=\bm 0,\dots,q_{v_{d(r,r-1)}}=0) \propto MLG(\bm 0, [\bm H_{\eta_c},\bm Q_{\eta_c}]^{-1},\bar{\bm\alpha}_{\eta_c},\bar{\bm\kappa}_{\eta_c})$

where $\bm H_{\eta_c}=[\bm\Psi'_c,\bm V_{\eta_c}']'$, $\bm Q_{\eta_c}=[\bm Q_{1\eta_c}',\bm Q_{2\eta_c}']'$, $\bar{\bm\alpha}_{\eta_c}=(\bm 1'_n,\bm\alpha'_{\eta_c})'$, $\bar{\bm\kappa}_{\eta_c}=((\bm t^{\bm\rho}\circ exp(\bm X_c\bm\beta_c+\bm\Psi_c\bm\eta+\bm\gamma_c))',\bm\kappa'_{\eta_c})'$.

3. The full-conditional distribution of $\bm\eta_d$ is:

$p(\bm\eta_d,\bm q_{\eta_d}|\bm t,\bm V_{\eta_d},\bm\alpha_{\eta_{d}},\bm\kappa_{\eta_{d}},\bm q_{\eta}=\bm 0,\dots,q_{v_{d(r,r-1)}}=0)\propto MLG(\bm 0, [\bm H_{\eta_d},\bm Q_{\eta_d}]^{-1}, \bar{\bm\alpha}_{\eta_d},\bar{\bm\kappa}_{\eta_d})$

where $\bm H_{\eta_d}=[\bm\Psi'_d,\bm V'_d]'$,$\bm Q_{\eta_d}=[\bm Q'_{1\eta_d},\bm Q'_{2\eta_d}]'$, $\bar{\bm\alpha}_{\eta_d}=(\bm Z'+\bm \zeta',\bm\alpha'_{\eta_d})'$, $\bar{\bm\kappa}_{\eta_d}=(exp(\bm X_d\bm\beta_d+\bm\Psi_d\bm\eta+\bm\gamma_d)',\bm\kappa'_{\eta_d})'$.

4. The full-conditional distribution of $\bm\beta_c$ is:

$p(\bm\beta_c, \bm q_{\beta_c}|\bm t,\bm\eta_c,\bm\alpha_{\beta_c},\bm\kappa_{\beta_c},\bm q_{\eta}=\bm 0,\dots,q_{v_{d(sj)}}=0)\propto MLG(\bm 0,[\bm H_{\beta_c},\bm Q_{\beta_c}]^{-1},\bar{\bm\alpha}_{\beta_c},\bar{\bm\kappa}_{\beta_c})$

where $\bm H_{\beta_c}=[\bm X'_c,\bm I'_{\beta_c}]'$, $\bm Q_{\beta_c}=[\bm Q_{1\beta_c}',\bm Q_{2\beta_c}']'$, $\bar{\bm\alpha}_{\beta_c}=(\bm 1'_{\beta_c},\bm\alpha'_{\beta_c})'$, $\bar{\bm\kappa}_{\beta_c}=((\bm t^{\bm\rho}\circ exp(\bm \Psi_c\bm\eta+\bm \Psi_d\bm\eta_c+\bm \gamma_c))',\bm\kappa'_{\beta_c})$.

5. The full-conditional distribution of $\bm\beta_d$ is:

$p(\bm\beta_d,\bm q_{\beta_d}|\bm Z,\bm\eta,\bm\eta_d,\bm\gamma_d,\bm\alpha_{\beta_d},\bm\kappa_{\beta_d},\bm q_{\eta}=\bm 0,\dots,q_{v_{d(sj)}}=0) \propto MLG(\bm 0, [\bm H_{\beta_d},\bm Q_{\beta_d}]^{-1},\bar{\bm\alpha}_{\beta_d},\bar{\bm\kappa}_{\beta_d})$

where $\bm H_{\beta_d}=[\bm X'_d,\bm I'_{\beta_d}]'$,$\bm Q_{\beta_d}=[\bm Q_{1\beta_d}',\bm Q_{2\beta_d}']'$, $\bar{\bm\alpha}_{\beta_d}=(\bm Z'+\bm\zeta',\bm\alpha'_{\beta_d})'$, $\bar{\bm\kappa}_{\beta_d}=(exp(\bm \Psi_d\bm\eta+\bm \Psi_d\bm\eta_d+\bm \gamma_d)'+\bm\zeta',\bm\kappa'_{\beta_d})'$.

6. The full-conditional distribution of $\bm\gamma_c$ is:
$p(\bm\gamma_c,\bm q_{\gamma_c}|\bm\alpha_{\gamma_c},\bm\kappa_{\gamma_c},\bm q_{\eta}=\bm 0,\dots,q_{v_(sj)}=0) \propto MLG(\bm 0,[\bm H_{\gamma_c},\bm Q_{\gamma_c}]^{-1},\bar{\bm\alpha}_{\gamma_c},\bar{\bm\kappa}_{\gamma_c})$

where $\bm H_{\gamma_c}=[\bm I'_{\gamma_c},\bm I'_{\gamma_c}]'$, $\bm Q_{\gamma_c}=[\bm Q'_{1\gamma_c},\bm Q'_{2\gamma_c}]'$, $\bar{\bm\alpha}_{\gamma_c}=(\bm 1'_{\gamma_c},\bm\alpha'_{\gamma_c})'$, $\bar{\bm\kappa}_{\gamma_c}=((\bm t^{\bm\rho}\circ exp(\bm X_c\bm\beta_c+\bm\Psi_c\bm\eta_c+\bm\Psi_c\bm\eta))', \bm\kappa'_{\gamma_c})'$.

7. The full-conditional distribution of $\bm\gamma_d$ is:
$p(\bm\gamma_d,\bm q_{\gamma_d}|\bm\alpha_{\gamma_d},\bm\kappa_{\gamma_d},\bm q_{\eta}=\bm 0,\dots, q_{v_{d(sj)}}=0) \propto MLG(\bm 0,[\bm H_{\gamma_d},\bm Q_{\gamma_d}]^{-1},\bar{\bm\alpha}_{\gamma_d},\bar{\bm\kappa}_{\gamma_d})$

where $\bm H_{\gamma_d}=[\bm I'_{\gamma_d},\bm I'_{\gamma_d}]'$, $\bm Q_{\gamma_d}=[\bm Q_{1\gamma_d}',\bm Q_{2\gamma_d}']'$, $\bar{\bm\alpha}_{\gamma_d}=(\bm Z'+\bm\zeta',\bm\alpha'_{\gamma_d})'$, $\bar{\bm\kappa}_{\gamma_d}=(exp(\bm X_d\bm\beta_d+\bm\Psi_d\bm\eta_d+\bm\Psi_d\bm\eta)'+\bm\zeta', \bm\kappa'_{\gamma_d})'$.

8. The full-conditional distribution of $V_{\eta(sj)}$ is:

$p(V_{\eta(sj)},q_{v_{\eta(j)}}|\bm\eta,\alpha_v,\kappa_v,\bm q_{\eta}=\bm 0,\dots,q_{v_{d(wk)}}=0) \propto MLG(\bm 0,\bm H_{v_{\eta(j)}},\bar{\bm\alpha}_{v_{\eta(j)}},\bar{\bm\kappa}_{v_{\eta(j)}})$

where $\bm H_{v_{\eta(j)}}=[\eta_j,1]'$, $\bm Q_{v_{\eta(j)}}=[Q_{1v_{\eta(j)}},Q_{2v_{\eta(j)}}]'$, $\bar{\bm\alpha}_{v_{\eta(j)}}=(\alpha_{\eta},\alpha_{v})'$, $\bar{\bm\kappa}_{v_{\eta(j)}}=(\kappa_{\eta},\kappa_{v})'$.

9. The full-conditional distribution of $V_{\eta_{c} (sj)}$ is:

$p(V_{\eta_c(sj)},q_{v_{\eta_c(j)}}|\bm\eta_c,\alpha_{v},\kappa_{v},\bm q_{\eta}=\bm 0,\dots,q_{v_{d(wk)}}=0)
\propto MLG(\bm 0,\bm H_{v_{\eta_c(j)}},\bar{\bm\alpha}_{v_{\eta_c(j)}},\bar{\bm\kappa}_{v_{\eta_c(j)}})$

where $s\neq w$ or $j\neq k$, $\bm H_{v_{\eta_c(j)}}=[\eta_{c(j)},1]'$, $\bm Q_{v_{\eta_c(j)}}=[Q_{1v_{\eta_c(j)}},Q_{2v_{\eta_c(j)}}]'$, $\bar{\bm\alpha}_{v_{\eta_c(j)}}=(\alpha_{v},\alpha_{v_{\eta_c(j)}})'$, $\bar{\bm\kappa}_{v_{\eta_c(j)}}=(\kappa_{\eta_c},\kappa_{v})'$.

10. The full-conditional distribution of $V_{\eta_{d} (sj)}$ is:

$p(V_{\eta_d(sj)},q_{v_{\eta_d (j)}}|\bm\eta_d,\alpha_{v},\kappa_{v},\bm q_{\eta}=\bm 0,\dots,q_{v_{c(wk)}}=0)
\propto  MLG(\bm 0,\bm H_{v_{\eta_d(j)}},\bar{\bm\alpha}_{v_{\eta_d(j)}},\bar{\bm\kappa}_{v_{\eta_d(j)}})$

where $s\neq w$ or $j\neq k$, $\bm H_{v_{\eta_d(j)}}=[\eta_{d(j)},1]'$, $\bm Q_{v_{\eta_d(j)}}=[Q_{1v_{\eta_d(j)}},Q_{2v_{\eta_d(j)}}]'$, $\bar{\bm\alpha}_{v_{\eta_d(j)}}=(\alpha_{\eta_d},\alpha_{v})'$, $\bar{\bm\kappa}_{v_{\eta_d(j)}}=(\kappa_{\eta_d},\kappa_{v})'$.

For the parameters in Parameter Model 8, the prior is:
\begin{equation*}
f(r_1,r_2,b)=exp\big\{r_1\alpha_a+r_2\kappa_a-b log(\Gamma(\alpha_a))-b \alpha_a log(\kappa_a)\big\},\ a=\eta, \eta_c,\eta_d, \beta_c,\beta_d,\gamma_c,\gamma_d
\end{equation*}
Then, the posterior for the parameters $(\alpha_a,\kappa_a)$ can be derived as follow (we use the $\bm\eta$ as an example):
\begin{align*}
& p(\alpha_{\eta},\kappa_{\eta}|\bm\eta,\bm V_{\eta}, r_1,r_2,b)\propto f(\bm\eta|\bm V_{\eta},\alpha_{\eta},\kappa_{\eta})f(\alpha_{\eta},\kappa_{\eta}|r_1,r_2,b)\\
\propto &  \Big(\frac{\kappa_{\eta}^{r\alpha_{\eta}}}{\Gamma(\alpha_{\eta})^r}\Big)exp\{\bm\alpha_{\eta}^{*'}\bm V_{\eta} \bm\eta-\bm\kappa_{\eta}^{*'}exp(\bm V_{\eta}\bm \eta)\}\times exp\big\{r_1\alpha_{\eta}+r_2\kappa_{\eta}-b log(\Gamma(\alpha_{\eta}))+b \alpha_{\eta} log(\kappa_{\eta})\big\}\\
\propto & exp\{\alpha_{\eta}\bm 1'_n\bm V \bm\eta-\kappa_{\eta}\bm 1'_n exp(\bm V\bm\eta)+r\alpha_{\eta}log(\kappa_{\eta})-rlog(\Gamma(\alpha_{\eta}))+r_1\alpha_{\eta}+r_2\kappa_{\eta}-b log(\Gamma(\alpha_{\eta}))+b \alpha_{\eta} log(\kappa_{\eta})\}\\
\propto & exp\big\{(\bm 1'_r\bm V\bm\eta+r_1)\alpha_{\eta}+(r_2-\bm 1'_rexp(\bm V\bm\eta))\kappa_{\eta}-(r+b)log(\Gamma(\alpha_{\eta}))+(r+b)\alpha_{\eta}log(\kappa_{\eta}) \big\}\\
\propto & exp\big\{r_{1\eta}\alpha_{\eta}+r_{2\eta}\kappa_{\eta}-b_{\eta}log(\Gamma(\alpha_{\eta}))+b_{\eta}\alpha_{\eta}log(\kappa_{\eta}) \big\}
\end{align*}
where $r_{1\eta}=\bm 1'_r\bm V\bm\eta+r_1$, $r_{2\eta}=r_2-\bm 1'_rexp(\bm V\bm\eta)$, $b_{\eta}=(r+b)$. According to \citet{bradley2015computationally}, the conditional distribution of $\alpha_{\eta}-1|\kappa_{\eta}$ is Conway-Maxwell-Poisson with parameter $exp({b_{\eta}}log(\kappa_{\eta})+r_{1\eta})$ and $b_{\eta}$, and one can use Taylor expansion to estimate the $exp({b_{\eta}}log(\kappa_{\eta})+r_{1\eta})$ in order to avoid computation difficulties with large dataset. 
The conditional distribution of $\kappa_{\eta}|\alpha_{\eta}$ is Gamma($\alpha_{\eta}b_{\eta}+1,-r_{2\eta}$). Therefore, in practice, we can use Conway-Maxwell-Poisson and Gamma distributions to update the parameters $(\alpha_{\eta}, \kappa_{\eta})$. Other shape and rate full-conditional distributions of hyper-parameters can be derived in similar way.

We choose $\alpha_v=1000$ and $\kappa_v=0.001$ for the prior on $\bm V$. This choice is motivated by the observation that large $\alpha$ and small $\kappa$, lead to a log-gamma distribution that approximately equals a standard multivariate Gaussian distribution\citep{bradley2018bayesian}.

\section{C: The Conditional MLG Distribution}
Let $\bm q =(\bm q_1',\bm q_2')'$, so that $\bm q_1$ is \textit{g}-dimensional and $\bm q_2$ is (\textit{m-g})-dimensional. Partition $\bm V^{-1}=[\bm{H \ B}]$ into an $m\times g$ matrix $\bm H$ and an $m\times$(\textit{m-g}) matrix $\bm B$ such that the inverse of $\bm V^{-1}$ exists. Then, $\bm q_1|\bm q_2=\bm{d,c,V,\alpha,\kappa}$ is called a conditional multivariate log-Gamma (cMLG) random vector with pdf 
\begin{equation}\label{cmlgpdf}
f(\bm q_1|\bm q_2=\bm{d,c,V,\alpha,\kappa})= \frac{1}{M} \ exp\big \{ \bm{\alpha}'\bm{Hq}_1-\bm{\kappa}'_{cond}exp(\bm{Hq}_1)\big\},
\end{equation}
where $M$ is the normalizing constant and $\bm{\kappa}'_{cond}\equiv (\kappa_{1,cond},\dots,\kappa_{m,cond})= exp\big\{\bm{Bd}-\bm V^{-1}\bm c+log(\bm\kappa)\big\}$. The density in (\ref{cmlgpdf}) is proportional to the full-conditional distribution in a Poisson/MLG hierarchical model. We use $cMLG(\bm c, \bm H, \bm \alpha, \bm \kappa)$ to represent a conditional multivariate log-Gamma distribution with those parameters above.

\section{D: Data Augmentation for conditional MLG Random Vectors}
As an example, consider we have a parameter $\bm y \sim cMLG(\bm c, \bm V^{-1}, \bm\alpha, \bm\kappa)$ with $\bm V^{-1}=[\bm H,\bm Q]$, where $\bm y$ is $r$ dimensional, $\bm H$ is $n \times r$, and $\bm Q$ is $n \times (n-r)$. Let $\bm c=-\bm V^{-1}\bm Q\bm q$, where $\bm q$ is a latent variable has the same length with $\bm y$, and $\bm Q$ is the orthogonal matrix in the QR decomposition of the matrix $\bm I-\bm H(\bm H'\bm H)^{-1}\bm H'$. Let $\bm q$ have the improper prior $f(\bm q)=1$. Then, the joint distribution of $\bm y$ and $\bm q$ is
\begin{align}\label{cgs}
f(\bm y,\bm q|\bm V,\bm\alpha,\bm\kappa)
&\propto MLG(\bm 0, [\bm H,\bm Q]^{-1}, \bm\alpha, \bm \kappa)\\
& \propto exp\big\{\bm\alpha'\bm H\bm y-\bm\alpha' \bm V\bm c-\bm \kappa' exp(\bm H\bm y-\bm V\bm c)\big\}\nonumber\\
&\propto exp\big\{\bm\alpha'\bm H\bm y+\bm\alpha'\bm Q\bm q-\bm\kappa'exp(\bm H\bm y+\bm Q\bm q)\big\}\nonumber\\
&\propto exp\big\{\bm\alpha'[\bm H, \bm Q](\bm y',\bm q')'-\bm\kappa'exp([\bm H, \bm Q](\bm y',\bm q')'\big\}\nonumber\\
&\propto MLG(\bm 0, [\bm H,\bm Q]^{-1}, \bm\alpha, \bm \kappa)
\end{align}

See Appendix for details. To simulate from (\ref{cgs}) we can use Equation (\ref{mlg}). That is, 
\begin{equation}\label{collapseSample}
\begin{pmatrix}\bm y\\ \bm q \end{pmatrix}
=\begin{pmatrix} 
(\bm H' \bm H)^{-1}\bm H' \bm w \\
\bm Q' \bm w
\end{pmatrix},
\end{equation}

Hence, we can collapse across $\bm q$ and obtain a simulated value from $f(\bm y|\bm V, \bm\alpha,\bm\kappa)$ rather easily using (\ref{collapseSample}). The argument in (\ref{cgs}) can be extended to the model in (\ref{wap}). However, this requires extensive book keeping. These details are provided in Appendix B.
\vfill
\bibliographystyle{biom}
\bibliography{JMLGM}

\begin{thebibliography}{}
\bibitem[\protect\citeauthoryear{Bradley, Wikle, and Holan}{Bradley
  et~al.}{2019}]{bradley2019hierarchical}
Bradley, J.~R., Wikle, C.~K., and Holan, S.~H. (2019).
\newblock Hierarchical models for spatial data with errors that are correlated
  with the latent process.
\newblock {\em Statistica Sinica} .
\end{thebibliography}
%\noalignite{*}

\end{document}